\documentclass[prl,superscriptaddress,twocolumn,showpacs]{revtex4}
\usepackage{graphicx}

\begin{document}

\title{Strongly Reshaped Organic-Metal Interfaces:
Tetracyanoethylene on Cu(100) }

\author{St\'ephane Bedwani}
\email{stephane.bedwani@polymtl.ca}
\affiliation{D\'epartement de g\'enie physique and Regroupement qu\'eb\'ecois sur les mat\'eriaux de pointe (RQMP), \'Ecole Polytechnique de Montr\'eal, Montr\'eal, Qu\'ebec H3C 3A7, Canada}

\author{Daniel Wegner}
\affiliation{Department of Physics,
University of California, Berkeley, and Materials Sciences
Division, Lawrence Berkeley National Laboratory, Berkeley,
California 94720-7300, USA}

\author{Michael F. Crommie}
\email{crommie@berkeley.edu}
\affiliation{Department of Physics,
University of California, Berkeley, and Materials Sciences
Division, Lawrence Berkeley National Laboratory, Berkeley,
California 94720-7300, USA}

\author{Alain Rochefort}
\email{alain.rochefort@polymtl.ca}
\affiliation{D\'epartement de g\'enie physique and Regroupement qu\'eb\'ecois sur les mat\'eriaux de pointe (RQMP), \'Ecole Polytechnique de Montr\'eal, Montr\'eal, Qu\'ebec H3C 3A7, Canada}


\begin{abstract}
The interaction of the strong electron-acceptor tetracyanoethylene
(TCNE) with the Cu(100) surface has been studied with scanning
tunneling microscopy experiments and first-principles density
functional theory calculations. We compare two different
adsorption models with the experimental results and show that the
molecular self-assembly is caused by a strong structural
modification of the Cu(100) surface rather than the formation of a
coordination network by diffusing Cu adatoms. Surface atoms become
highly buckled and the chemisorption of TCNE is accompanied by a
partial charge-transfer.

\end{abstract}
\maketitle

Molecules containing multiple cyano (CN) groups such as
tetracyanoethylene (TCNE) \citep{khatkale},
7,7,8,8-tetra\-cyanoquinodimethane (TCNQ)~\citep{kamma} or
2,3,5,6-tetra\-fluoro-7,7,8,8 tetra\-cyano\-quinodimethane
(F4-TCNQ)~\citep{romaner,jackel} represent the archetype of strong
electron acceptors. Furthermore, there is a promising development
in the design of molecule-based magnets with high Curie
temperatures that are based on strongly charged ligands anchored
to paramagnetic transition metal
atoms~\citep{manriquez,tengstedt,Jain,Harvey}. However, a
significant amount of structural characterization still needs to
be carried out in order to understand and describe the interaction
of such strong acceptor ligands with metallic substrates. Recent
studies on the adsorption of TCNE and F4-TCNQ on single crystal
surfaces such as Cu(111)~\citep{erley,romaner} and
Au(111)~\citep{jackel,wegner} reveal the existence of strongly
bound species where a large variation in charge density is
invariably observed near the interface. In a comparative study of
TCNE adsorption on various metal surfaces it was found that TCNE
strongly self-assembles into chains on the Cu(100) surface.
Moreover, scanning tunneling microscopy (STM) images contain
additional bright protrusions and dark trench features that have
been interpreted as consequences of a strongly reconstructed
surface with buckled surface Cu atoms~\citep{wegner}.

In this Letter, we present results of first-principles density functional theory (DFT)
calculations and STM simulations as well as STM experiments of the
structural and electronic properties of adsorbed TCNE on Cu(100).
We found that TCNE is strongly adsorbed on the surface, and the
adsorption gives rise to a reconstruction of the surface
where the topmost Cu atoms bonded to TCNE are highly buckled
with respect to the remaining surface Cu atoms. We have identified
the presence of resonance peaks below the Fermi level that are
associated to the highest occupied molecular orbital (HOMO) and to the lowest unoccupied molecular orbital (LUMO) of TCNE.


Electronic structure calculations were carried out using DFT within a local density
approximation (LDA) \citep{LDA} included in the \textsc{siesta}
package \citep{SIESTA}. For core electrons, norm-conserving
pseudopotentials following a Troullier and Martins
scheme~\citep{TM} were used with relativistic corrections for Cu
atoms. For the representation of valence electrons, we used an
extended atomic basis set of polarized double-$\zeta$ type. The
one-dimensional TCNE chain was built periodically from a supercell
containing one molecule and a Cu(100) slab of three atomic layers
with about 3$\times$8 Cu atoms in each layer. All atoms were
allowed to relax except for Cu atoms in the bottom layer.
Geometries were fully optimized using a Broyden scheme until a
tolerance of 10$^{-3}$~Ry/Bohr was reached for each system under
study along with their deformation and adsorption energy, total
density of states (DOS), projected DOS (PDOS), Mulliken population
analysis, and charge density. STM simulations were performed with
our \textsc{spags-stm} (Strongly Parallel Adaptive Grid Solvers --
STM) software to evaluate topographic mode images and scanning
tunneling spectra (STS). The software includes several algorithmic
strategies such as parallel computation of the tunnel
currents~\citep{STM_GREEN} and adaptive grids that minimize the
probing sites needed to obtain a high resolution
image~\citep{MAMI}. In STM simulations, the tunnel current was
computed within a scattering approach based on the
Landauer-B\"uttiker formalism~\citep{buttiker01} along with an
extended H\"uckel theory (EHT) Hamiltonian~\citep{STM_GREEN,cerda01}. The
sample preparation of TCNE/Cu(100) and STM experiments (using a
Pt/Ir tip) were performed as described in
Ref.~\onlinecite{wegner}.

\begin{figure}
\begin{centering}
\includegraphics[scale=0.85]{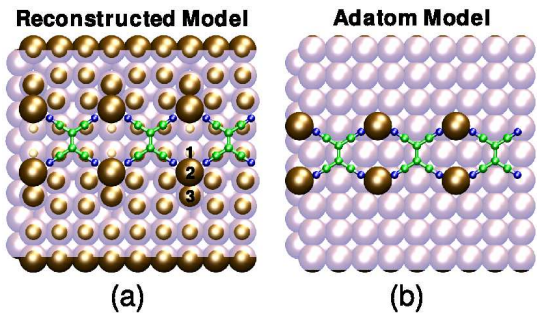}
\par\end{centering}
\caption{\label{fig:model} DFT-LDA optimized structural geometry
for the one-dimensional chain arrangement of TCNE on the Cu(100)
surface (a) for the reconstructed model and (b) for the adatom model. A transparent plane is used to emphasize the magnitude of reconstruction of the topmost Cu layer.}
\end{figure}

As a starting point, we intuitively placed the TCNE molecule in a
highly symmetric adsorption site on a flat Cu(100) surface. As
shown in Fig.~\ref{fig:model}(a), the DFT-LDA geometry
optimization reveals that a TCNE chain assembly induces an strong
reconstruction of the topmost Cu(100) surface layer, most
dramatically where the Cu-N bonds are formed. Within the
one-dimensional chain structure, we can identify three different
Cu atoms that are involved in the surface reshaping process: one
Cu atom underneath the molecule chain is pushed below the surface
plane by 0.3~{\AA} (marked as ``1''), another surface atom (marked
as ``2'') is highly buckled by 1.3~{\AA} out of the surface plane,
and finally a third surface atom outside the chain (``3'') opens a
trench by sliding laterally by 1.0~{\AA} toward the chain and
0.3~{\AA} out of the surface plane. The distance between highly
buckled Cu atoms is 7.7~{\AA} along the chain direction and
6.6~{\AA} in the perpendicular direction. The energy difference
between the buckled and a flat Cu(100) surface is 3.55~eV, which
gives a deformation energy of 1.78~eV per buckling. Strong
interaction of adsorbates with substrates can induce nanoscale
surface reshaping~\citep{LargeMol_rev}. For example, a single
molecule C$_{90}$H$_{98}$ at the step edge of a Cu(110) surface
was used to build a molecule-sized electrode~\citep{Rosei01}. In
another case, the removal of C$_{60}$H$_{66}$ self-assembly from a
Cu(110) surface left imprinted trenches underneath~\citep{C60H66}.
Metal surfaces roughening with C$_{60}$ were also
reported~\citep{C60_Au110,C60_Pd110,C60_Cu110_Ni110}.

On this reconstructed surface, TCNE is strongly bonded to Cu atoms. The energy needed to dissociate the TCNE/Cu(100) complex into individual but unrelaxed components is 6.60~eV. Adsorbed TCNE
is significantly deformed with respect to the gas phase structure, the relaxation of the molecule gives a deformation energy of 0.71~eV. Hence, the formation of this strongly buckled structure would necessitate a minimal energy amount of (3.55~eV + 0.71~eV) = 4.26~eV, which is still 2.34~eV lower than the calculated
dissociation energy for the TCNE/Cu(100) complex. Although energies can be more accurately obtained by considering GGA functionals and basis set superposition error (BSSE) \citep{hobbs}, the comparison of relative values obtained at LDA level should give an appropriate description. In addition, the deformation of isolated TCNE contributes to reduce its HOMO/LUMO gap from 2.61~eV to 1.98~eV.

While the DFT-LDA optimization of the TCNE chain on Cu(100)
suggests the described strong surface reconstruction, an entirely
different self-assembly mechanism of molecules on Cu(100) is known
to exist due to the fact that Cu adatoms from step edges can
easily diffuse along the Cu(100) terraces~\citep{boisvert}. These
Cu adatoms can connect to molecules and promote ordered molecular
structures via formation of a coordination
network~\citep{lin,tait}. In order to consider such a
self-assembly process for the TCNE chains on Cu(100), we also
performed DFT-LDA calculations on the adsorption of TCNE with four
Cu adatoms located in four-fold sites of the Cu(100) surface. The
calculated optimized geometry is shown in Fig.~\ref{fig:model}(b).
In this case, the surface layer remains unreconstructed, and the
adatoms have left their four-fold positions to establish covalent
bonding with CN groups. The surface structure of this adatom model
shows a distance between Cu adatoms of 7.7~{\AA} along and
5.4~{\AA} across the chain, the latter being significantly shorter
(by 1.2~{\AA}) than in the reconstructed model. In addition, the
dissociation energy for this complex is 7.67~eV, which makes the
TCNE more tightly bound to the surface than in the reconstructed
model.

\begin{figure}
\begin{centering}
\includegraphics[scale=0.95]{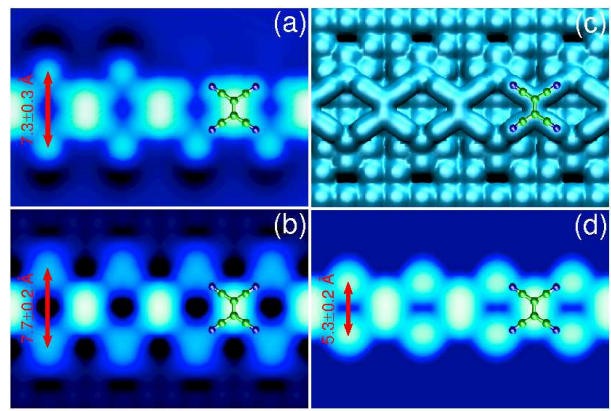}
\par\end{centering}
\caption{\label{fig:STM}Topographic STM images of a TCNE chain (a)~obtained
experimentally with a Pt/Ir tip (V=1~mV, I=5~nA) from Ref.~\onlinecite{wegner}, and (b)~calculated on a
reconstructed Cu(100) surface with a Pt(111) tip (V=1~mV, I=15~nA). (c)~Calculated
charge density of the valence electrons for the reconstructed
model (a black plane is used to highlight the density drop).
(d)~Calculated STM image of a TCNE chain for the adatom model. All image domains are 31~{\AA}$~\times$~18~{\AA}.}
\end{figure}

Although the exclusion of one of the two mechanisms remains a
difficult task without investigating the dynamics of the
processes, we can try to discriminate between them by comparing
how well simulated STM images of both models resemble the
experimental STM images of the TCNE chains, which show three
characteristic features (Fig.~\ref{fig:STM}(a)): (\emph{i})~The
TCNE molecules appear as a bright, elongated protrusion centered
around the C=C bond and four faint lobes close to the CN groups
(referred to as \emph{short legs} in Ref.~\onlinecite{wegner}).
(\emph{ii})~Additional protrusions next to CN groups of
neighboring molecules above and below the TCNE chain
(\emph{extended legs}) are presumably caused either by buckled Cu
surface atoms or by bound Cu adatoms~\footnote{The absence of a few
extended legs ($\sim10\%$) indicates that the reconstruction or
the diffusion of adatoms is a thermally activated process that has
not reached complete equilibrium. 
The presence of defects, although rarely observed here, may also contribute to the creation of missing legs}. (\emph{iii})~Dark trenches
surround the buckled atoms or adatoms above and below the TCNE
chain.

The STM simulation of the reconstructed model probed with a
Pt(111) tip is displayed in Fig.~\ref{fig:STM}(b) \footnote{Simulated STM images using an Ir(111) tip lead to the same results.}. The theoretical STM image can reproduce the first
two features, although the \emph{short} and \emph{extended legs}
appear more blurred. The third feature, the dark trench, is faint
but visible in the STM simulation. Moreover, we observe a clear
drop in the calculated charge density map close to the buckled
atoms (Fig.~\ref{fig:STM}(c)), which can be associated to the dark
trench area. Overall, the characteristic experimental features are
reproduced fairly well. In comparison, the simulated STM image of the adatom model is
shown in Fig.~\ref{fig:STM}(d). While also this model
qualitatively reproduces the bright TCNE center as well as the
\emph{short} and \emph{extended leg} features (that also appear
blurred), no dark trench can be reproduced at all. Hence, a main
characteristic feature of the experimental images is missing in
this model.

Finally, we quantitatively compare the features related to the
\emph{extended legs} in experimental and simulated STM images (see
arrows in Fig.~\ref{fig:STM}). From the experiments we find
distances of 7.7$\pm$0.1~{\AA} along the chain and 7.3$\pm$0.3~{\AA} across. Both
theoretical models show a distance of 7.7$\pm$0.2~{\AA} along the chain
(i.e., three times the Cu nearest-neighbor distance), which is in
very good agreement with the experiment. Across the TCNE chain,
the situation changes: for the reconstructed model, the simulated
STM image shows a distance of 7.7$\pm$0.2~{\AA}, which is slightly larger
than the actual distance of the highly buckled atoms, because the
Cu atom ``3'' next to the trench has a significant
contribution to the apparent height in the STM image. The
simulated STM image of the adatom model exhibits a distance of
only 5.3$\pm$0.2~{\AA} across the chain, similar to the actual adatom
distance across the TCNE chain. This is in strong disagreement to
the experiment. Consequently, the reconstructed model reproduces
the experimental observations far better, and thus we conclude
that the strong buckling of the Cu(100) surface is most likely the
driving mechanism for the observed self-assembly of TCNE molecules
into chains. In addition, the DFT and STM calculations
agree to support that mobile adatoms do not apparently participate in the formation of the one-dimensional
TCNE chain: the most stable geometry is not experimentally
observed by STM and the STM simulations on the reconstructed model
are accurately reproducing the experimental STM results.

Another important aspect of the interaction of TCNE with surfaces
is the magnitude of charge transfer. The formation of TCNE$^{1-}$
or TCNE$^{2-}$ species is generally assumed when low work function
metals such as Na or Ca are used~\citep{khatkale}. The situation
is slightly different with transition metals for which the
accumulation of net charge on TCNE depends on the magnitude of
TCNE$\rightarrow$metal donation but also on the
metal$\rightarrow$TCNE backdonation~\cite{romaner}. Hence, a
fractional net charge on TCNE would reflect that a charge
transferred to the metal (donation) is significantly returned
back to the molecule (backdonation).

\begin{figure}
\begin{centering}
\includegraphics[scale=0.32]{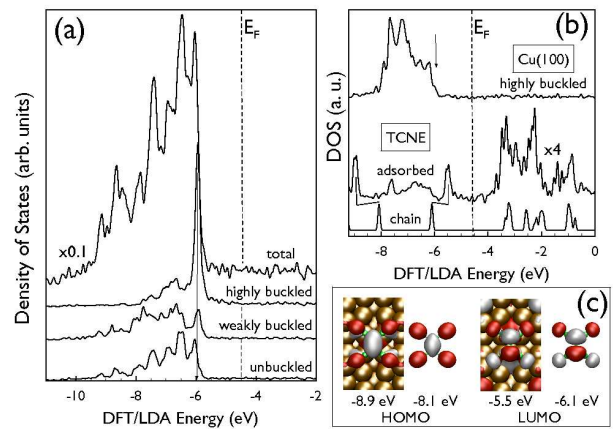}
\par\end{centering}
\caption{(a)~Total DOS and projected DOS for a buckled Cu surface
without TCNE and (b)~after adsorption of TCNE, and total DOS of an isolated TCNE one-dimensional chain. (c)~Wavefunction contours of the molecular orbitals located at $-8.9$~eV and $-5.5$~eV in (b) and HOMO and LUMO of isolated TCNE.
\label{fig:DOS}}
\end{figure}
Fig.~\ref{fig:DOS}(a) shows the total DOS and various PDOS for a
buckled surface without the TCNE molecule. The total DOS and PDOS
over most Cu atoms of the buckled surface are similar to the DOS
diagram usually observed for this Cu surface~\citep{dBands},
except for highly buckled Cu atoms for which the states are more
strongly localized (sharp and intense) and more weakly bonded than
other Cu surface states. The adsorption of TCNE is considered in
Fig.~\ref{fig:DOS}(b): the sharp peak of highly buckled Cu atoms
at $-6.0$~eV is now smeared out within the broad PDOS peaks at
higher binding energy. The disappearance of the localized surface
states into the higher binding energy manifold suggests an
electron depletion of buckled Cu atoms and a strong mixing of
states with TCNE orbitals. In Fig.~\ref{fig:DOS}(b), the DOS of an
isolated chain shows quite narrow peaks while the PDOS of adsorbed
TCNE is much broader. Therefore, such broadening results from the
important mixing between TCNE and Cu(100) states rather than from
the coupling between TCNE molecules in the chain.

The nature of two TCNE-related molecular orbitals near the Fermi
level are shown in Fig.~\ref{fig:DOS}(c). We can easily attribute
the peaks at $-8.9$~eV and $-5.5$~eV of adsorbed TCNE to the HOMO
($-8.1$~eV) and LUMO ($-6.1$~eV) of isolated TCNE, respectively.
The shifts with respect to the isolated TCNE chain can be
explained in terms of charge transfer: the HOMO moves toward high
binding energies (electron donation to Cu) and the LUMO moves
toward low binding energies (electron backdonation to TCNE). 
The contribution from TCNE to the wavefunction of the TCNE/Cu(100) complex centered at $-5.5$~eV
is $\sim$90\%, i.e. TCNE accepts at least $1.8$~$|e|$ from Cu(100). 
Mulliken population analysis of the TCNE/Cu(100) system
gives a net charge of $0.30$~$|e|$ on TCNE, which indicates that Cu 
receives $\sim$$1.5$~$|e|$ from HOMO and HOMO-$n$ states of TCNE. A
comparable charge transfer of $0.55$~$|e|$ from a polycrystalline
Cu surface to TCNE has been deduced from surface-enhanced Raman
scattering experiments~\citep{SERS}. The adsorption of the
electron acceptor F4-TCNQ on the Cu(111) surface also leaves a net
charge on the molecule of approximately
$0.6$~$|e|$~\citep{romaner}.

In summary, slab DFT-LDA calculations, STM experiments and
simulations reveal that the adsorption of TCNE molecules on
Cu(100) is accompanied by a reconstruction on the topmost Cu
surface layer. By comparison, we exclude the possibility of an
alternative self-assembly mechanism based on diffusing Cu adatoms.
Reconstructing Cu atoms that bind directly to the CN groups of
TCNE are strongly buckled out of the surface plane by 1.3~{\AA}.
The presence of dark trenches in STM experiments and simulations
is attributed to a drop in the surface charge density that is
caused by the concerted lateral displacement of Cu atoms
contributing to the buckling process. TCNE interacts strongly with
the Cu(100) surface leading to a large adsorption energy. However,
the net charge transfer onto the TCNE moiety is relatively weak
due to a significant amount of charge that is backdonated to the
molecule.

\begin{acknowledgments}
This work was supported by the Natural Sciences and Engineering
Research Council of Canada (NSERC), and by the US National Science
Foundation through NSF NIRT Grant ECS-0609469. We are also
grateful to the RQCHP for providing computational facilities. S.B. is grateful
to the Fonds qu\'eb\'ecois de la recherche sur la nature et les
technologies (FQRNT) for scholarship. D.W. thanks the Alexander
von Humboldt Foundation for a research fellowship. We also thank
Ryan Yamachika for technical assistance and stimulating
discussions.
\end{acknowledgments}

\end{document}